\documentclass{article}

\usepackage{PRIMEarxiv}

\usepackage[utf8]{inputenc} 
\usepackage[T1]{fontenc}    
\usepackage{hyperref}       
\usepackage{url}            
\usepackage{booktabs}       
\usepackage{amsfonts}       
\usepackage{nicefrac}       
\usepackage{microtype}      
\usepackage{lipsum}
\usepackage{fancyhdr}       
\usepackage{graphicx}       
\usepackage{amsmath}
\graphicspath{{media/}}     
\title{Simulation-Informed Deep Learning for Enhanced SWOT Observations of Fine-Scale Ocean Dynamics}
\author{
  E. Cutolo\textsuperscript{1,2}, 
  C. Granero-Belinchon\textsuperscript{1,2},
  P. Thiraux\textsuperscript{1,2}, 
  J. Wang\textsuperscript{3}, 
  R. Fablet\textsuperscript{1,2} \\
  \textsuperscript{1}IMT Atlantique, Lab-STICC, UMR 6285, 29238, CNRS, Brest, France \\
  \textsuperscript{2}ODYSSEY Team-Project, INRIA Ifremer IMT-Atl., 35042, CNRS, Brest, France \\
  \textsuperscript{3}Jet Propulsion Laboratory, California Institute of Technology, Pasadena, CA, USA \\
  \texttt{eugenio.cutolo@imt-atlantique.fr},
}

\begin{document}
\maketitle

\begin{abstract}
Oceanic processes at fine scales are crucial yet difficult to observe accurately due to limitations in satellite and in-situ measurements. The Surface Water and Ocean Topography (SWOT) mission provides high-resolution Sea Surface Height (SSH) data, though noise patterns often obscure fine-scale structures. Current methods struggle with noisy data or require extensive supervised training, limiting their effectiveness on real-world observations.

We introduce SIMPGEN (Simulation-Informed Metric and Prior for Generative Ensemble Networks), an unsupervised adversarial learning framework combining real SWOT observations with simulated reference data. SIMPGEN leverages wavelet-informed neural metrics to distinguish noisy from clean fields, guiding realistic SSH reconstructions. Applied to SWOT data, SIMPGEN effectively removes noise, preserving fine-scale features better than existing neural methods. This robust, unsupervised approach not only improves SWOT SSH data interpretation but also demonstrates strong potential for broader oceanographic applications, including data assimilation and super-resolution.
\end{abstract}

\keywords{Satellite altimetry \and Deep learning \and SWOT \and KaRIn noise \and Submesoscale ocean dynamics}

\section{Introduction}
Oceanic processes at scales of 5–150 km, evolving over days to weeks, fundamentally influence the ocean’s physical and ecological state by redistributing heat and nutrients, driving water mass formation, and modulating air–sea exchanges \cite{Ferrari2009OceanSinks, Callies2013InterpretingKm, Siegelman2019EnhancedFronts}. Their energetic nature also shapes broader circulation patterns, underscoring the need to observe and model submesoscale phenomena \cite{Klein2019Ocean-ScaleSpace}.

However, capturing fine-scale features remains challenging. In situ platforms (e.g., moorings, Argo floats, gliders) provide detailed but sparse snapshots \cite{Chelton2001ChapterAltimetry}, while nadir along-track satellite sensors balance resolution, accuracy, and revisit frequency, often limiting their ability to observe small-scale variability \cite{Morrow2019GlobalMission, Ballarotta2019OnMaps}. The Surface Water and Ocean Topography (SWOT) mission and its  Ka-band Radar Interferometer (KaRIn) \cite{Morrow2019GlobalMission, Fu2024TheWater} provide unprecedented high-resolution SSH observations. However, they involve complex noise processes \cite{Treboutte2023KaRInProducts}, potentially obscuring small-scale SSH signals. 

High-resolution numerical modeling \cite{Xu2022OnOcean} and data assimilation \cite{Carrassi2018DataPerspectives} fill observational gaps by integrating measurements into ocean models, yet remain prone to biases from sub-grid parameterizations, limited and irregularly-sampled data and observation noise \cite{Chelton2001ChapterAltimetry, Siegelman2019EnhancedFronts}. Deep learning (DL) offers a complementary solution for estimating ocean variables from incomplete observations \cite{Arcucci2021DeepAssimilation, Fablet2021End-to-endCurrents, Nardelli2022Super-ResolvingAlgorithms, Febvre2024TrainingData,Cutolo2024CLOINet:Learning}. Networks trained on high-resolution simulations can predict quantities like subsurface temperature \cite{Zhang2023DerivingModel}, sea surface height (SSH) \cite{Martin2023SynthesizingAnomalies, Febvre2022JOINTMODELS}, or velocity fields \cite{Fablet2023MultimodalSynergies}, but often struggle when confronted with sparse, noisy, and real-world data \cite{Morrow2019GlobalMission, Buizza2022DataLearning}. Domain adaptation methods help but require extensive, well-characterized observational datasets \cite{Febvre2024TrainingData, Tzachor2023DigitalSustainability}. As a result, the operational SWOT SSH denoising  \cite{Treboutte2023KaRInProducts} based on a supervised learning from simulation data cannot fully reveal the potential of SWOT SSH observations, especially regarding fine-scale patterns. 

Unsupervised learning provides an additional pathway for exploiting real measurements lacking ground-truth labels \cite{Zhao2024ApplicationsReview}. Among these methods, Generative Adversarial Networks (GANs) \cite{Goodfellow2014GenerativeNetworks} have been used in oceanography for gap filling and super-resolution \cite{Wang2022STA-GAN:Data, Izumi2022Super-resolutionMethods, Nardelli2022Super-ResolvingAlgorithms}. Building on these ideas, we introduce an adversarial learning framework for unsupervised processing of SWOT SSH observations referred to as SIMPGEN (Simulation Informed Metric and Prior for Generative Ensemble Network). Unlike standard GAN approaches, our method jointly learns: (i) a neural metric space for comparing oceanographic fields, and (ii) a neural inversion operator that synthesizes physically consistent fields from partial data.


SIMPGEN pairs a neural inversion operator generating noise-free, dynamically plausible reconstructions with an adversarial discriminator that enforces realism. We show that SIMPGEN mitigates SWOT KaRIn noise and preserves submesoscale structures more effectively than operational SWOT denoising \cite{Treboutte2023KaRInProducts}.

The paper is organized as follows: \autoref{section:data} presents the SWOT mission, KaRIn noise characteristics, and the synthetic and real datasets. \autoref{section:methods} details the SIMPGEN framework. \autoref{section:results} illustrates its performance in reconstructing SWOT SSH fields. Finally, \autoref{section:conclusion} discusses implications for ocean state estimation and potential synergies with data assimilation.

\section{Context and Case study: SWOT mission and the KaRIn noise}
\label{section:data}
Launched in December 2022 by NASA, CNES, the Canadian Space Agency, and the UK Space Agency, the Surface Water and Ocean Topography (SWOT) mission addresses the need for higher-resolution observations of global ocean dynamics \cite{Morrow2019GlobalMission, Fu2024TheWater}. Its Ka-band Radar Interferometer (KaRIn) delivers sea surface height (SSH) measurements at a wavelength resolution of about 10–15 km far exceeding the ~150 km resolution of previous altimeters \cite{Beauchamp20234DVarNet-SSH:Altimetry,Archer2020IncreasingAltimetry}. This improvement provides a valuable window into submesoscale processes and strongly nonlinear internal waves that shape ocean kinetic energy and circulation.

However, KaRIn data remain prone to wave-modulated noise, particularly at finer spatial scales (e.g., 2 km), limiting accurate estimates of geostrophic velocities, vorticity, and strain rates. Standard filtering (e.g., median, Lanczos) may oversmooth small-scale features, and direct supervised training of machine learning models is hampered by the lack of noiseless ground truth \cite{Treboutte2023KaRInProducts}.

Two experiments are performed, one on simulated and the other on real SWOT data. In both cases, we use the eNATL60 model \cite{Ajayi2020SpatialModels} as reference, which resolves the North Atlantic with a grid spacing of about 1.7 km. For purely simulation-based experiment, synthetic SWOT observations are generated by adding KaRIn-like noise scaled by significant wave height and distance from nadir to the model’s true SSH fields. Although actual SWOT noise has proven to be lower than prelaunch estimates by roughly a factor of four \cite{Chelton2024AVorticity}, the synthetic data remain valuable for benchmarking denoising methods.

For the real-world experiment, we replace the noisy observations with actual SWOT Level-2 KaRIn Low Rate SSH data. These real observations were collected during SWOT's initial "fast-sampling" phase a roughly three-month period when the satellite operated in a 1-day repeat orbit, providing frequent observations on a swath-aligned 2 km grid to capture rapid SSH variability \cite{Wang2022OnTopography}. We benchmark our proposed method against the state-of-the-art CLS denoising technique, which employs a similar U-Net architecture but follows a supervised training approach.

\begin{figure}[ht] \centering \includegraphics[width=1.0\textwidth]{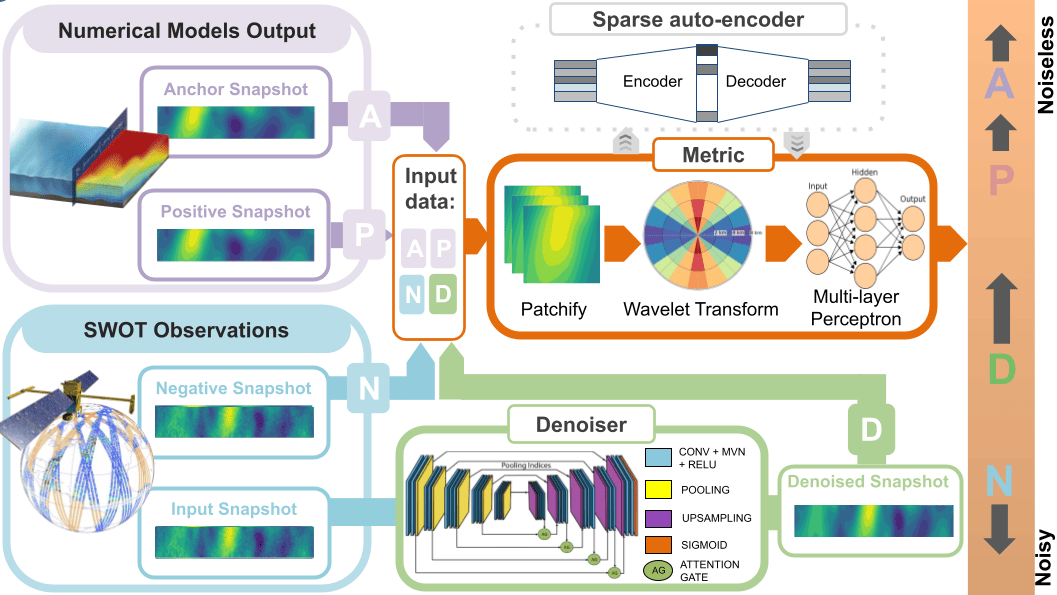} \caption{Schematic of the proposed SIMPGEN architecture.} \label{fig:architecture} 
\end{figure}

\section{Simulation-Informed Metric and Prior for Generative Ensemble Network}
\label{section:methods}

We introduce an original unsupervised deep learning scheme to reconstruct physically-consistent ocean fields from noisy observations by leveraging numerical simulations as reference data (see \autoref{fig:architecture}). This approach builds upon adversarial learning, a methodology made popular by GAN. Unlike traditional supervised learning, which requires paired datasets of observations and true states, adversarial methods operate in an unsupervised fashion. Specifically, our approach relies on two unpaired datasets: \(\mathcal{O}\), a collection of noisy observations, and \(\mathcal{S}\), a reference dataset of numerical simulations. The reference dataset is assumed to approximate the distribution of true but inaccessible ocean states that correspond to the observations in \(\mathcal{O}\).

We draw inspiration from variational data assimilation (DA) schemes, which are among the state-of-the-art methods in operational oceanography to solve inverse problems from observation datasets~\cite{Carrassi2018DataPerspectives}. Formally, we state the reconstruction of the state \(\mathbf{x}\) as the resolution of a minimization problem according to a variational cost function that balances prior information with observational data. This cost function of the state \(\mathbf{x}\)  is expressed as:
\begin{equation}
\label{eq:da_cost}
J(\mathbf{x}; \mathbf{y}) = J_{\text{prior}}(\mathbf{x}) + \frac{1}{2} (\mathbf{H}(\mathbf{x}) - \mathbf{y})^\top \mathbf{R}^{-1} (\mathbf{H}(\mathbf{x}) - \mathbf{y}),
\end{equation}
where \(\mathbf{y}\) refers to the observations, \(\mathbf{H}\) to the observation operator mapping the model state into observation space, and \(\mathbf{R}\) to the observation error covariance matrix under the assumption of an additive Gaussian noise. The term \(J_{\text{prior}}(\mathbf{x})\) encodes the prior knowledge about the state. In the widely used 3D-Var approach \cite{Cutolo2022DiagnosingApproach}, the prior relies on Gaussian assumptions and is defined as:
\begin{equation}
\label{eq:gaussian_prior}
J_{\text{prior}}(\mathbf{x}) = \frac{1}{2} (\mathbf{x} - \mathbf{x}_b)^\top \mathbf{B}^{-1} (\mathbf{x} - \mathbf{x}_b),
\end{equation}
where \(\mathbf{x}_b\) is referred to as the background state, and \(\mathbf{B}\) is the background error covariance matrix. This formulation results in solving a classic linear-quadratic minimization problem. However, the assumption of a Gaussian prior in \(\mathbf{B}\) cannot represent the complex and non-linear dynamics exhibited by ocean states, especially for the smaller mesoscale processes. 

To address these shortcomings, we introduce a novel prior term defined as the Mean Squared Error (MSE) in a latent space between the denoised observation and an ensemble of states $\mathbf{x}_s$ from the reference dataset \(\mathcal{S}\) as follows:

\begin{equation}
\label{eq:metric_space_prior}
J_{\text{prior}} (\mathbf{x}) = \mathbb{E}_{\mathbf{x}_s \in \mathcal{S}} 
\left[ \| M_{\Phi}(\mathbf{x}) - M_{\Phi}(\mathbf{x}_s) \|^2 \right].
\end{equation}
where \(M_\Phi\) is a trainable neural mapping between the ocean state space and a latent space. It defines a neural metric space to asses whether a given state $\mathbf{x}$ lies in the distribution associated with some reference dataset. We may point that \autoref{eq:gaussian_prior} can be regarded as a specific case of \autoref{eq:metric_space_prior}, where the mapping operator $M_{\Phi}(\mathbf{x})$ involves the projection of the state on the square-root of the inverse of the covariance $\mathbf{B}$. Computationally,  the expectation in \autoref{eq:metric_space_prior} is approximated through an empirical mean over samples from the reference simulation dataset 
\(\mathcal{S}\).

To solve the resulting variational problem, rather than considering an iterative gradient-based descent of \autoref{eq:da_cost}, we leverage a neural inversion operator $G_{\theta}$, with trainable parameters $\theta$. This operator maps observations \(\mathbf{y}\) and the observation error covariance $\mathbf{R}$ to the state estimates $\hat{x}$. The variational cost of \autoref{eq:da_cost} defines the training loss of $G_{\theta}$ such that:
\begin{equation}
\label{eq:neural_operator_training}
\widehat{\theta} = \arg \min_{\theta} \mathbb{E}_{\mathbf{y} \in \mathcal{O}} \left[ J\bigl(\widehat{\mathbf{x}}, \mathbf{y}\bigr) \right] \mbox{ subject to } \widehat{\mathbf{x}} = G_\theta(\mathbf{y},\mathbf{R})
\end{equation}
This training formulation implicitly assumes that the mapping operator \(M_\Phi\) is known. As detailed in \autoref{section:adversarial_learning}, we explore a joint training of both the neural metric space defined by the mapping operator \(M_\Phi\), and the neural inversion operator $G_{\theta}$ using an adversarial learning strategy. Whereas the latter is analogous to the generator in GAN framework, the neural metric space relates to the definition of a discriminator, such that reference ocean states shall be associated to low values of the prior (\autoref{eq:metric_space_prior}). The adversarial learning then involves a competition between the two operators to deliver after convergence a discriminative latent space and an efficient inversion operator. More details on these Neural Network operators can be found in the supplementary material.

\subsection{Adversarial learning scheme}
\label{section:adversarial_learning}

Following a standard adversarial learning scheme \cite{Goodfellow2014GenerativeNetworks}, the training phase alternates optimization steps for the parameters of the neural inversion operator \(G_{\theta}\) according to variational cost (\autoref{eq:neural_operator_training}) and for the parameters of the neural metric space defined by mapping operator \(M_{\Phi}\).  
The primary objective in designing \(M_{\Phi}\) is to preserve the structure of the reference dataset \(\mathcal{S}\), which approximates the true distribution of ocean states, while efficiently distinguishing states outside it. To achieve this, \(M_{\Phi}\) maps states into a latent space optimized for meaningful distance computations, formalized by the following minimization problem:
\begin{equation}
\label{eq:phi_optimization}
\widehat{\Phi} = \arg \min_{\Phi} \mathbb{E}_{(\mathbf{x}_a, \mathbf{x}_p) \in \mathcal{S}, \mathbf{x}_n \sim G_\theta} \left[ J_{\text{adv}}(M_\Phi(\mathbf{x}_a), M_\Phi(\mathbf{x}_p), M_\Phi(\mathbf{x}_n)) \right],
\end{equation}
where \(J_{\text{adv}}\) is the adversarial cost function defined as a triplet loss:
\begin{equation}
\label{eq:triplet_loss}
J_{\text{adv}}(M_\Phi(\mathbf{x}_a), M_\Phi(\mathbf{x}_p), M_\Phi(\mathbf{x}_n)) = 
\left[ \|M_\Phi(\mathbf{x}_a) - M_\Phi(\mathbf{x}_p)\|^2 - \|M_\Phi(\mathbf{x}_a) - M_\Phi(\mathbf{x}_n)\|^2 + \text{margin} \right]_+,
\end{equation}
The hinge function \([\,\cdot\,]_+\) ensures a non-negative loss. In this setup, \(\mathbf{x}_a\) (anchor) and \(\mathbf{x}_p\) (positive) are physically-consistent states sampled from the reference dataset \(\mathcal{S}\), representing realistic ocean states. Conversely, \(\mathbf{x}_n\) (negative) is a state generated by the fixed neural inversion operator \(G_\theta\) from noisy observations. This part of the adversarial framework focuses solely on enhancing the discriminative power of the neural metric space \(M_\Phi\), ensuring it cannot be misled by an irrealistic generated outputs of \(G_\theta\). The margin hyperparameter ensures that anchor-positive pairs remain closer in the latent space than anchor-negative pairs, sharpening \(M_\Phi\)’s ability to evaluate the consistency of ocean state estimates. In summary, the adversarial learning procedure alternates the following two steps until convergence:
\begin{enumerate}
    \item \textbf{Update \(\theta\)}: Minimize the expected DA cost function in \autoref{eq:neural_operator_training} with fixed \(\Phi\), updating \(\theta\) to improve the generative operator \(G_{\theta}\).
    \item \textbf{Update \(\Phi\)}: Minimize the triplet loss in \autoref{eq:phi_optimization} with fixed \(\theta\), updating \(\Phi\) to refine the metric operator \(M_{\Phi}\).
\end{enumerate}
This alternating scheme ensures a balance between improving the neural inversion and refining the latent space where meaningful distances are preserved.

By combining the variational assimilation formulation with a adversarial learning approach, the proposed framework  allows the inversion operator \(G_{\theta}\) to produce plausible state estimates from observations, while the metric operator \(M_{\Phi}\) ensures these estimates reside in a latent space that accurately captures the structure of physically-consistent states. This approach effectively bypasses the need for groundtruthed training datasets in standard supervised training schemes, while making the most of available yet unpaired observation and simulation datasets.   

\subsection{SWOT Denoising Training}
\label{section:swot_denoising_training}
The proposed framework is well-suited for reconstructing SSH fields from the noisy observations provided by SWOT. While SWOT offers unprecedented spatial resolution, its observations are inherently noisy, obscuring small-scale dynamics critical for oceanographic studies.

We conducted two separate experiments and thus trainings: one using synthetic noisy observations \(\mathcal{O}\), derived by applying synthetic KARIN noise to high-resolution SSH fields from the numerical model ENATL60, and another using real SWOT observations as \(\mathcal{O}\). In both cases, the synthetic SSH fields from ENATL60 without noise serve as the reference dataset \(\mathcal{S}\). 

For this specific aim, the adversarial training alternatively work with \autoref{eq:neural_operator_training} and \autoref{eq:triplet_loss}, where \(\mathbf{y}\) represents the noisy SSH from \(\mathcal{O}\). A slight modification compared to the generic case in \autoref{section:adversarial_learning} is that the minimization involves \(\mathbf{x}_n\), sampled from both the purely noisy SSH fields and the outputs of \(G_{\theta}\), ensuring a better convergence.

The neural inversion operator \(G_{\theta}\) is designed to handle SWOT-specific data characteristics, including its unique swath geometry and nadir gap. To reflect the uncertainty associated with KARIN noise, random fluctuations were incorporated into \(\mathbf{R}\) during training, modeled as a Gaussian random variable with a mean representing the nominal observation error (defining the global weight) and a variance set to 10\% of this value (capturing relative uncertainty). At inference, we leverage an ensemble-driven strategy: we sample \(\mathbf{R}\)  repeatedly to generate 30 denoised realizations, then average them for the final output. This ensemble approach broadens the range of noise scenarios addressed, enhancing robustness and yielding more reliable SSH estimates.

The neural metric space operator \(M_{\Phi}\) learns to assess the consistency of reconstructed SSH fields relative to the statistical properties of \(\mathcal{S}\). Using a wavelet-based approach (see the supplementary material), \(M_{\Phi}\) captures the multi-scale and directional dynamics of SSH fields, which are crucial for ensuring physically meaningful reconstructions. For more details on both of the operators please refer to the supplementary material.

\section{Results and Discussion}
\label{section:results}
Here, we present the results to illustrate the effectiveness of the proposed approach for denoising SWOT observations. As previously mentioned, we considered two different and independent scenarios: one using purely simulated data, where both the reference dataset and the noisy dataset (with added KaRIn-like noise) were available, and another combining simulated data as a reference with real SWOT observations.

In both cases, to analyze the directional distribution of energy, we performed two distinct averaging processes: one over snapshots where the along-track direction dominated and another where the across-track direction dominated. This separation was achieved using a simple masking criterion, where snapshots were classified based on whether the wavelet coefficient in the across-track direction exceeded that in the along-track direction. This approach ensures that directional differences are preserved rather than averaged out, allowing for a more detailed assessment of the noise structure and the effectiveness of the denoising method. For a detailed explanation of how the polar plots of wavelet coefficients correspond to the SSH fields, please refer to the supplementary material.

\subsection{Synthetic SWOT Observations}
We first consider an application to the SWOT denoising benchmark based on synthetic noisy SWOT observations derived from high-resolution numerical simulations. In this controlled setting, where the ground truth is available, we compare our method against the state-of-the-art baseline, a U-Net trained in a supervised manner. Specifically, the U-Net is provided with paired noisy and true SSH fields during training. In contrast, our adversarial learning scheme is trained without direct supervision, accessing only the noisy fields alongside the uncoupled reference SSH. The used dataset consists of the entire data-challenge SWOT orbital cycle 13, which includes 270 passes. In particular, we used the left swath for training and tested on the right swath after flipping it to preserve the symmetric nadir-distance-dependent error.

\begin{figure}[ht]
    \centering
    \includegraphics[width=\textwidth]{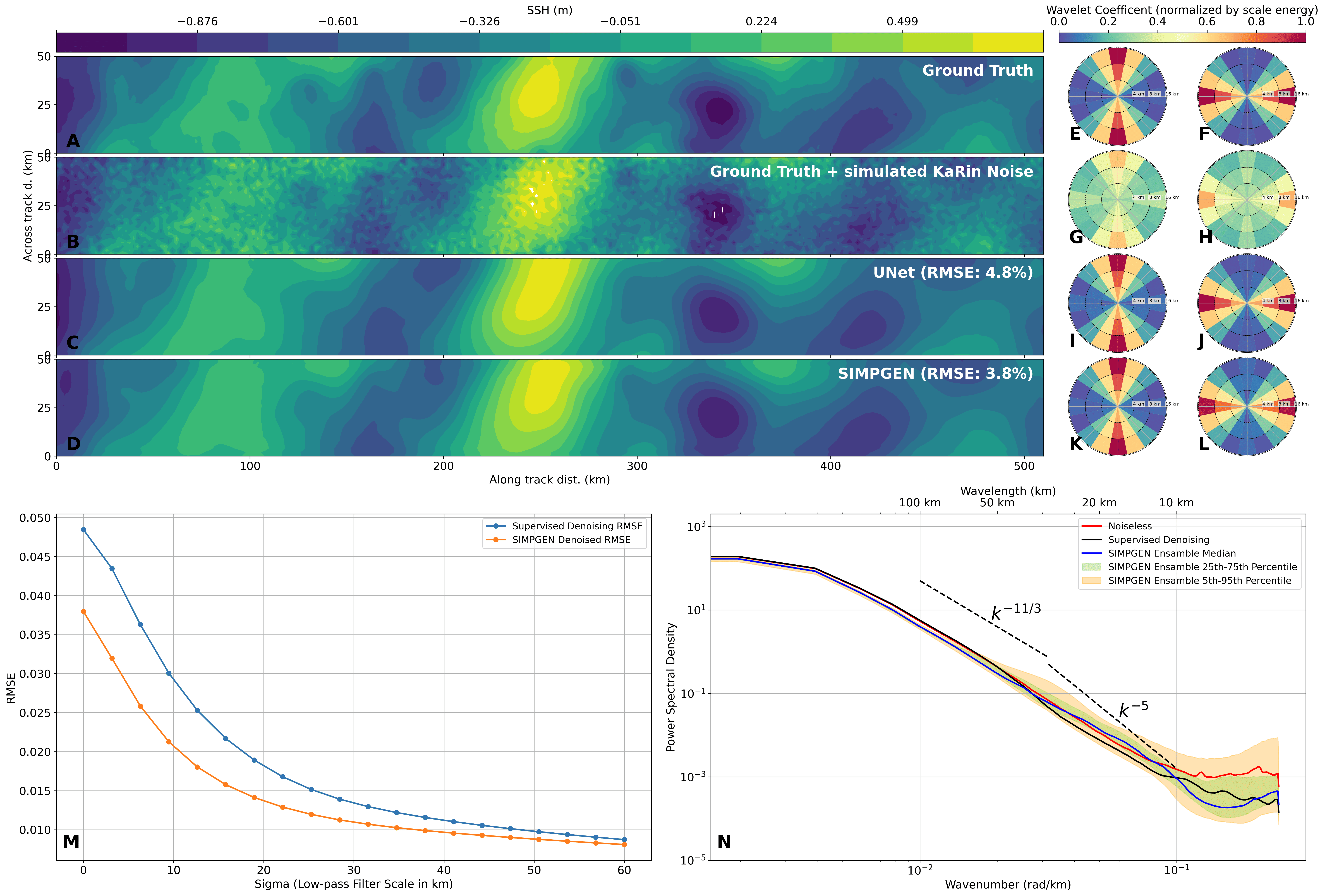}
    \caption{\textit{(A–D)} Example snapshot of SSH fields: (A) noiseless ground truth, (B) noisy field, (C) supervised U-Net denoised field, and (D) SIMPGEN denoised field. \textit{(E–L)} Polar plots of the average wavelet coefficients normalized by the scale energy for two cases: across-track dominated snapshots (right column) and along-track dominated snapshots (left column). Panels (E,F) represent noiseless fields, (G, H) noisy fields, (I, J) supervised U-Net denoised fields, and (K, L) SIMPGEN denoised fields. \textit{(M)} RMSE as a function of spatial scale, where the neural inversion operator outperforms the supervised model, particularly at smaller scales. \textit{(N)} Power spectral density comparison across wavelengths.}
    \label{fig:denoising_rmse}
\end{figure}

\autoref{fig:denoising_rmse} provides a comprehensive visual comparison between the ground truth (Panel A), the noisy dataset (Panel B), and the reconstructions obtained using both the baseline U-Net (Panel C) and our method (Panel D). While the supervised U-Net produces a smoother representation, the neural inversion operator more effectively preserves small-scale structures, particularly in regions with high spatial gradients. The RMSE averaged over the entire test dataset (Panel M) further supports this advantage: our trained neural inversion operator consistently outperforms the baseline U-Net denoiser, particularly when preserving all scales. In contrast, when smaller scales are smoothed out, the RMSE difference between the two methods becomes less pronounced.

The results suggest that training the U-Net with an RMSE-based loss leads to excessive smoothing, whereas our method effectively balances noise removal and detail preservation by leveraging the adversarial learning framework. Naturally, more sophisticated training strategies for the baseline such as incorporating additional terms in the loss function could improve performance. However, such modifications would inevitably increase training complexity and the risk of overfitting to the training dataset.

Panels E–L present the average wavelet coefficients for the different datasets, normalized by the maximum scale energy, for an along-track-dominated average (left column) and an across-track-dominated average (right column). The noiseless reference (Panels E, F) exhibits coherent energy distributions aligned with the dominant spatial directions. Conversely, in the noisy cases (Panels G, H), the wavelet coefficients display a significant spread of energy across multiple directions and scales, revealing the impact of the simulated KaRIn noise. Although the simulated KaRIn noise has an inherent anisotropy due to its dependence on cross-swath distance  and the significant wave height (SWH), we observe that, on average, this anisotropy does not manifest clearly in the wavelet coefficients.

Both the baseline U-Net denoiser (Panels I, J) and SIMPGEN (Panels K, L) yield nearly indistinguishable results in restoring the original anisotropy, demonstrating comparable effectiveness in noise reduction while maintaining the expected energy distributions.

In the average along-track spectra shown in Panel N, we show that SIMPGEN achieves a spectral energy distribution that preserves small-scale wavelengths, closely aligning with the noiseless reference. By contrast, the baseline U-Net denoiser slightly underestimates small-scale spectral energy, in agreement with the oversmoothing effect discussed earlier.

In summary, our neural inversion operator effectively removes isotropic noise while preserving anisotropy associated with small-scale features and maintaining the physical coherence of the SSH field across dominant scales and directions. Notably, our method achieves these results without relying on a supervised dataset that in the following experiment, the real SWOT observations, is not available.

\subsection{Demonstration with Real SWOT Observations}
\label{sec:others}
\begin{figure}[hp]
    \centering
    \includegraphics[height=0.8\textheight]{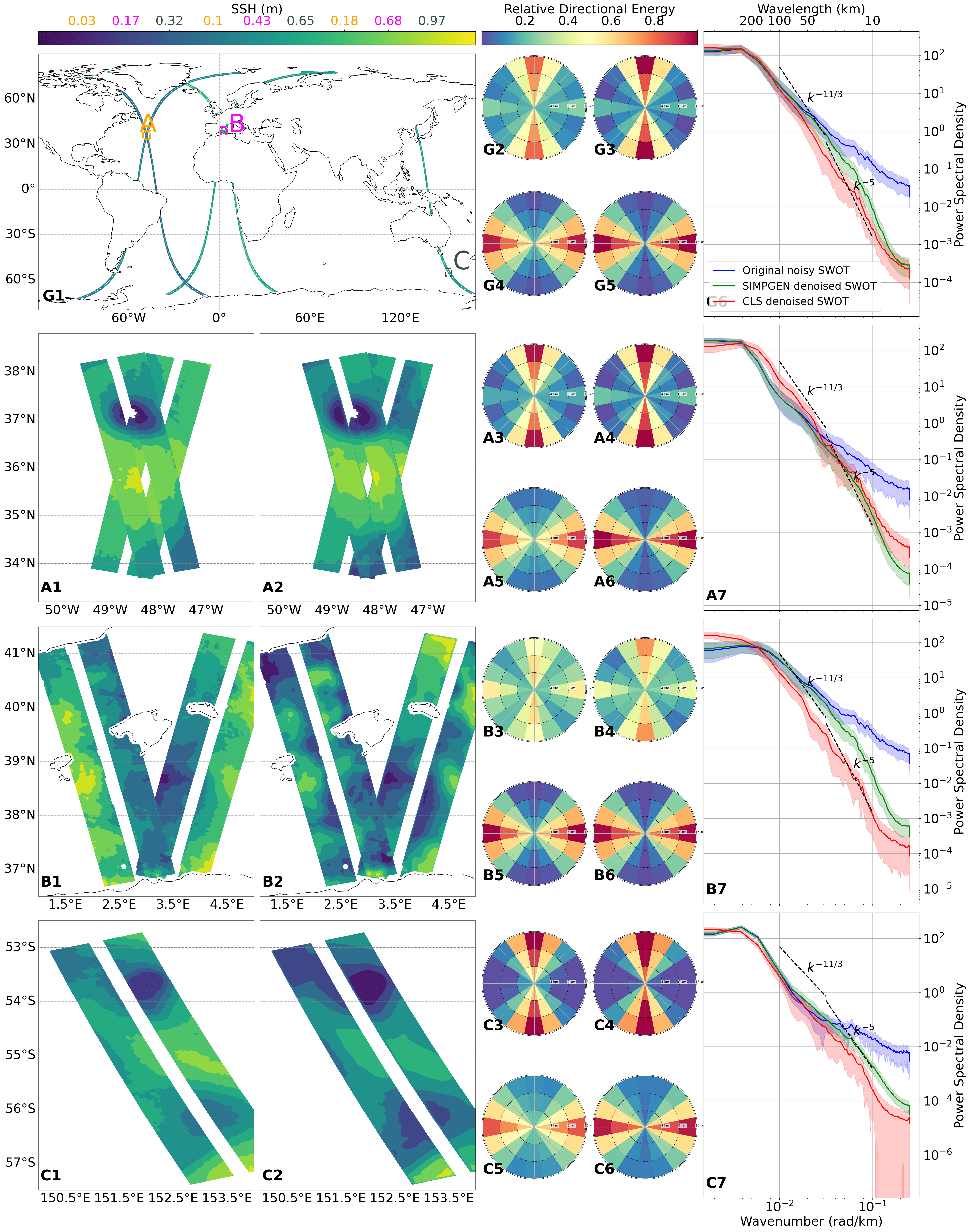}
    \caption{The figure presents the global (G1–G6) and regional analyses (Gulf Stream: A1–A7, Mediterranean Sea: B1–B7, Southern Ocean: C1–C7) in a structured panel layout. The first panel of the global analysis (G1) highlights the SWOT tracks and regions of interest, examples of SSH snapshots are provided in the regional rows. We use a single colorbar, but each region has its own numeric range, indicated by color coded ticks corresponding to the region box. For all analyses, the subsequent panels compare noisy fields (A1, B1, C1) with SIMPGEN denoised fields (A2, B2, C2) through the following elements: example snapshots of SSH fields, relative directional energy derived from wavelet coefficients normalized by the maximum scale energy (panels 3,4,5 and 6), and averaged across-track power spectral density (7). The wavelet energy analysis includes two datasets: along-track-dominated averages in panels 3 and 4, and across-track-dominated averages in panels 5 and 6. The panels 7 display the averaged power spectral density, where blue (red) curves represent the original (denoised) results, and shaded areas indicate 24-hour average variability for each scale. Dashed lines show constant spectral slopes, serving as references for typical dynamic regimes.}
\label{fig:spectral_density_comparison}
\end{figure}

The denoising of real SWOT SSH observations was evaluated in three distinct oceanic regions: the Mediterranean Sea, the Gulf Stream in the Atlantic Ocean and the Southern Ocean. SWOT ground tracks are highlighted in Panel G1 of~\autoref{fig:spectral_density_comparison}. As described in \autoref{section:data}, we focused on the fast-sampling phase, during which SWOT revisited each region approximately every 24 hours for at least 90 days. Since no ground truth is available for real SWOT observations, we used the same clean eNATL60 simulation from the synthetic experiment as a reference dataset for the unsupervised training.

Given that the directional wavelet analysis is performed on 50 km × 50 km patches, we expect to observe anisotropy depending on the dominant direction selected (along-track or across-track). This is because SSH fields reflect the dominance of physical processes, which typically exhibit coherent structures with directional continuity, such as mesoscale eddies and fronts. In contrast at such spatial domain, isotropy is generally associated with noise at small scales, as noise tends to distribute energy more uniformly across scales and directions without preserving physical coherence.

\autoref{fig:spectral_density_comparison} panels G2–G5 illustrate wavelet coefficients averaged along-track (G2 and G3) and across-track (G4 and G5) over the full dataset (across all three regions) of noisy (G2 and G4) and denoised (G3 and G5) fields. The overall effect of denoising is an increase in energy along the dominant direction in both the along-track and across-track averages. Specifically, energy is reduced in the direction orthogonal to the dominant one, while small-scale energy increases in alignment with the dominant direction (4 km coefficient).

To indirectly assess the effectiveness of the denoising process, we leveraged the 24-hour revisit cycle to evaluate the temporal consistency of the energy spectra. Additionally, we compared the resulting spectra with those obtained from the state-of-the-art Level-3 denoising provided by CLS \cite{Treboutte2023KaRInProducts}. As shown in Panel G6, our method yields spectra with improved temporal coherence compared to both the CLS method and the raw SWOT data. Furthermore, our method preserves more energy within the spatial scales spanning 50–10 km. For a more visual comparison with the CLS product please refer to the supplementary material.

Beyond this global perspective, a regional analysis provides insights into how our denoising technique applies to different oceanic regions with different dynamical characteristics. Since KaRIn error includes random instrument noise and geophysical induced errors, evaluating the denoising impact within different regional contexts allows us to assess how well the method preserves real ocean variability. For each region, we present example SSH snapshots, analyses of wavelet energy distribution, and power spectral density (PSD) averaged over all available data.

\subsubsection{Atlantic Ocean – Gulf Stream Region (Panels A1–A7)}
The Gulf Stream region exhibits energetic mesoscale eddies and strong frontal gradients leading to larger signal-to-noise ratio. In the noisy SSH field (Panel A1), we observe high-frequency noise contaminating mesoscale features, which is significantly reduced after denoising (Panel A2).

Panels A3–A6 illustrate wavelet energy distributions of noisy fields (A3 and A5) and after denoising (A4 and A6), highlighting the suppression of small-scale noise while retaining the dominant mesoscale structures. The 24-hour PSD variation (Panel A7) confirms this improvement, showing a significant reduction in high-frequency noise while preserving mesoscale variability. Notably, in this region the CLS denoising exhibits an energy distribution most similar to SIMPGEN. We hypothesize that this similarity arises because the supervised denoising method was trained on synthetic data specifically generated in this region. However, the CLS denoising still excessively enhances larger scales and smooths out part of the signal at smaller scales.

\subsubsection{Mediterranean Sea (Panels B1–B7)}
The Mediterranean Sea is characterized by relatively low-energy ocean dynamics. In the noisy SSH field (Panel B1) and its denoised counterpart (Panel B2), we observe submesoscale features that are of particular interest.

The wavelet energy distributions reveal that both the noisy (B3 and B5) and denoised (B4 and B6) fields exhibit relatively isotropic energy distributions, particularly in the cross-track averages (B5 and B6). This is consistent with the small first baroclinic Rossby radius of deformation in the Mediterranean (5–15 km), which is significantly smaller than in the open ocean (30–50 km). As a result, mesoscale eddies and filaments tend to be more compact and isotropic, meaning their structures are less elongated compared to those in the Gulf Stream.

When analyzing 50 km × 50 km patches, as done here, these small-scale structures are better sampled in all directions, reducing the dominant anisotropic patterns observed in other regions. Additionally, in this area, the across-track direction aligns more with large-scale flow structures, making anisotropy more evident in across-track dominated snapshots.

The 24-hour spectral variability (Panel B7) shows a clear reduction in high-frequency energy, indicating effective noise suppression. Notably here, the original SSH variability in the noisy dataset is slightly lower than in the other regions, likely due to the generally smaller significant wave height (SWH) in the Mediterranean. In this region, the CLS denoising notably underestimates the regional energy distribution, possibly due to the fact that the dynamics of its training region differ significantly from those of the Mediterranean.

\subsubsection{Southern Ocean (Panels C1–C7)}
The Southern Ocean is highly energetic, influenced by strong westerly winds and intense mesoscale eddy activity generated by the Antarctic Circumpolar Current (ACC). The noisy SSH field (Panel C1) and its denoised counterpart (Panel C2) reveal that denoising restores sharper mesoscale structures, particularly enhancing the definition of frontal features and eddy boundaries.

Panels C3–C6 display the wavelet energy distributions, while Panel C7 demonstrates that denoising effectively reduces small-scale temporal variability, indicating successful noise suppression. Notably, here the spectra at scales under 10 km is flatter than in the other regions, further explaining the significant temporal variability observed in the CLS denoising, which struggles to capture the region's dynamics accurately.

The larger SSH gradients observed in the Southern Ocean are likely due to the strong meridional gradients imposed by the ACC and its meandering structure. Unlike other regions, along-track SSH structures in the Southern Ocean exhibit strong anisotropy, likely reflecting the ACC meanders.

We observe that the strong SSH gradients of this region are less influenced by noise in the along-track direction, as C3 (noisy) and C4 (denoised) are quite similar. However, we still observe a suppression of small-scale energy in the orthogonal direction, which is then redistributed into more coherent smaller-scale structures aligned with the dominant flow. This is an indirect sign that the SWOT noise is mostly affecting the 4 km scale.

\section{Conclusion} 
\label{section:conclusion}
In this study, we introduced SIMPGEN (Simulation-Informed Metric and Prior for Generative Ensemble Networks), a deep-learning framework designed for the unsupervised reconstruction of physically consistent oceanographic fields from noisy and sparse observational data. By leveraging adversarial learning guided by simulation-based priors, SIMPGEN directly exploits information from realistic numerical models and actual observational datasets, without requiring explicit noisy-clean training pairs a key advantage when reliable ground-truth references are unavailable. While our framework is broadly applicable to various ocean variables and observational contexts, here we demonstrated its efficacy specifically on Sea Surface Height (SSH) fields measured by the SWOT mission, highlighting its capability to robustly mitigate complex, KaRIn noise and recover fine-scale oceanographic structures.

Our adversarial framework trains a neural metric that quantifies the similarity between observed fields and reference oceanographic data by analyzing scale- and direction-dependent energy distributions in the wavelet domain. By distinguishing physically plausible fields from noise-contaminated ones, this metric offers an interpretable measure of data fidelity, supporting both quality assessment and uncertainty quantification. Its ability to capture scale- and direction-specific information is especially advantageous for handling KaRIn noise, which is spatially structured, scale-dependent, and lacks direct ground-truth references in real SWOT observations—posing significant challenges to conventional denoising methods. Consequently, current denoising approaches rely on synthetic datasets to circumvent this lack of ground-truth data. In this work, we used these simulated SWOT observations to provide a controlled test environment for evaluating our approach. Under these conditions, SIMPGEN consistently outperforms a supervised U-Net baseline in preserving small-scale oceanographic features, achieving lower RMSE values and more faithful spectral energy distributions. By contrast, the supervised U-Net tends to oversmooth SSH fields and underestimate high-frequency energy, thus diminishing fine-scale variability.

Notably, real SWOT measurements can exhibit lower noise levels than those predicted by simulations under comparable sea states, suggesting even greater denoising potential for high-wave conditions (e.g., sea states above 6 meters). By learning from realistic simulations and seamlessly adapting to observational noise, SIMPGEN addresses the crucial gap encountered when ground-truth data are unavailable. Building on these findings, we tested our method on real SWOT observations in diverse oceanic regions including the Mediterranean Sea, the Gulf Stream, and the Southern Ocean to demonstrate its adaptability to region-specific noise characteristics. Directional wavelet analysis confirms that denoising significantly reduces noise while preserving coherent oceanic structures, particularly in snapshots dominated by across-track energy, where KaRIn noise is most pronounced. The 24-hour revisit cycle analysis further underscores temporal consistency, with SIMPGEN-processed fields retaining more coherent SSH signals across consecutive SWOT passes. Moreover, comparisons with a state-of-the-art SWOT denoiser \cite{Treboutte2023KaRInProducts} favor SIMPGEN in terms of stronger adherence to the original signal on larger spatial scales, reduced temporal variability at 24-hour intervals, and better alignment with the expected theoretical spectral slope.

Despite these promising results, several considerations remain. The wavenumber spectra of denoised fields typically exhibit steeper slopes (around $-5$) consistent with Charney geostrophic turbulence, yet this steepness could partially be an artifact, implying the model may inadvertently remove genuine ocean signals (e.g., internal waves with shallower spectral slopes). These observations counsel caution in interpreting denoised data and underscore the importance of accounting for unresolved oceanic processes when applying generative denoising methods. Finally, to promote deeper investigation of the physical impacts of our approach, we will release a publicly accessible denoised SWOT dataset alongside this paper, enabling broader evaluation and exploration of SIMPGEN’s capabilities.

Looking ahead, an exciting direction is to expand SIMPGEN to other oceanic variables such as sea surface temperature or velocity fields and to integrate it into operational data assimilation systems for more accurate ocean state estimation. Unlike many supervised approaches that must be retrained for each new instrument or noise characteristic, SIMPGEN’s generality across diverse observational settings hints at its potential to evolve into a foundation model for oceanography, one that can assess the realism of varied oceanic fields and power future digital twins of the ocean. This unsupervised, simulation-informed framework further sets the stage for a wide range of applications in emerging high-resolution satellite missions and other data scarce scenarios, making SIMPGEN a promising tool for refining satellite-based ocean data and ultimately enhancing our understanding of fine-scale ocean dynamics.

\section*{Acknowledgments}
This work was supported by the French National Research Agency (ANR-21-CE46-0011-01), within the program Appel à projets générique 2021. This work was supported by the Bryttany region gouvernment (22006858), within the program Stratégie d’Attractivité Durable 2022. This work was supported by the Finistere department gouvernment, within the program Aide aux Projets de REcherche 2022. J. Wang is supported by the NASA Physical Oceanography Program grant 80NSSC25K7643. E. Cutolo would like to extend his sincerest gratitude to S. Bolumar, whose unwavering support, patience, and encouragement were invaluable throughout the course of this work. 

\bibliographystyle{unsrt}  
\bibliography{references_sync}  

\newpage

\begin{appendix}

\section{Wavelet decomposition and analysis of synthetic fields}
The advent of 2D observational capabilities provided by SWOT (Surface Water and Ocean Topography) has greatly enhanced our ability to capture ocean surface dynamics. Traditional spectral analyses, which usually rely on azimuthally averaged energy spectra, may overlook critical directional structures in oceanic fields. Directional wavelet analysis addresses this limitation by providing a multi-scale, multi-orientation decomposition of oceanographic signals, thereby revealing energy anisotropy at various scales and improving our understanding of ocean surface processes.

The wavelet transform decomposes a given field \( f(x,y) \) into coefficients obtained by convolving \( f \) with a family of wavelet functions that are localized in both space and scale. In two dimensions, wavelets possess directional selectivity and can detect anisotropic features such as elongated structures, edges, or directional gradients often found in geophysical and oceanographic data. The transform at scale \( s \) and orientation \( \theta \) is given by
\[
W_f(s, \theta, x, y) 
= \int \int f(x', y') \,\psi_{s, \theta}(x - x', y - y')\,dx'\,dy',
\]
where \( \psi_{s, \theta} \) is a scaled and rotated mother wavelet function. Varying \( s \) and \( \theta \) produces a complete multi-scale, multi-directional view of the field.

In Kymatio’s library implementation, the default wavelets are based on a Morlet-type construction that is scaled and rotated according to the number of scales \( J \) and orientations \( L \). Specifically:
\begin{itemize}
\item {Number of scales, \( J \): wavelets are created at dyadic scales \( 2^j \) for \( j = 0, 1, \dots, J-1 \). Each wavelet is obtained by dilating the mother Morlet wavelet.}
  
\item {Number of orientations, \( L \): For each scale, \( L \) rotated versions of the wavelet are used, each rotated by a multiple of \( \pi / L \). This provides directional selectivity across \( L \) equally spaced orientations over \([0, \pi)\).}
\end{itemize}

Thus, for every \( j \in \{0,\dots,J-1\} \) and \( l \in \{0,\dots,L-1\} \), we construct a wavelet \(\psi_{j,l}\) by:
\begin{itemize}

\item {Scaling the mother Morlet wavelet by a factor \(s=2^j\).}
\item {Rotating it by the angle \(\theta_l = \tfrac{\pi l}{L}\).}
\end{itemize}

Consequently, for each pair \((j, l)\), you obtain a wavelet that captures features of a specific size \(\sim 2^j\) and orientation \(\theta_l\), allowing a multi-scale, multi-directional analysis of the input signal or image.

The Fourier transform decomposes signals into global sinusoidal components and loses spatial localization, whereas wavelets allow multi-scale analysis while preserving coherence in space. Spatial averaging of wavelet coefficients does not destroy the underlying directional energy distribution, an essential aspect for capturing anisotropic processes like oceanic turbulence and internal waves. In contrast, standard Fourier methods, which decompose the signal into globally defined sinusoidal components, can make it more challenging to isolate or interpret anisotropic structures without specialized filtering or directional decompositions.

\begin{figure}[hp]
    \centering
    \includegraphics[height=0.8\textheight]{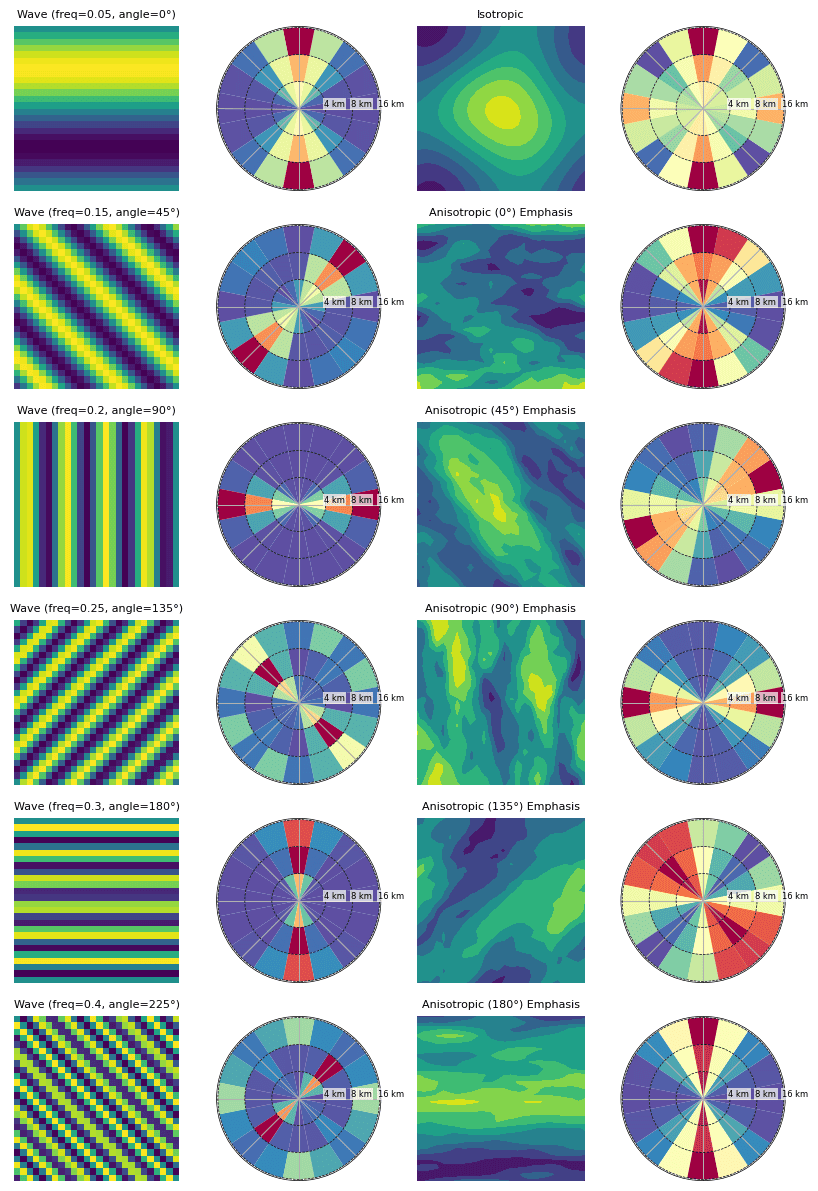}
    \caption{Applications of directional wavelet analysis on synthetic plane-wave fields and anisotropic power-law correlated fields. The first column shows synthetic plane-wave fields at various orientations and wavelengths and the second column presents the corresponding wavelet coefficients, where the radial and angular coordinates represent scale and orientation, respectively. The third column displays power-law fields with an energy spectrum following \(k^{-11/3}\), and the fourth column shows the wavelet analysis of these fields, revealing isotropic and anisotropic structures across scales.}
    \label{fig:spectral_density_comparison2}
\end{figure}

In \autoref{fig:spectral_density_comparison2}, we perform a wavelet analysis on: synthetic plane-wave fields and synthetic fields with power-law correlations (spectra following \(k^{-11/3}\)). In both cases, fields with different preferential orientations and wavelengths are used to illustrate how wavelet analysis detects directional energy concentrations and, in the case of power-law fields, how it  showcases both isotropic and anisotropic structures across scales. By revealing how energy is distributed as a function of orientation and scale, directional wavelet methods constitute a powerful tool for analyzing complex 2D SWOT observations that may exhibit strong anisotropy and spatial variability.

\section{Neural inversion operator} 
\label{section:neural_inversion_operator}

Regarding the neural inversion operator $G_{\theta}$, we consider a U-Net architecture widely recognized for its effectiveness in image-to-image translation tasks \cite{Weng2015U-Net:Segmentation}. It involves a multi-scale encoder-decoder scheme between the inputs, here noisy ocean observations and their error covariance, and the output, a denoised ocean field. Through skip connections between encoder and decoder layers, the network preserves multi-scale features and retains fine spatial details after denoising. We also include attention gates \cite{Oktay2018AttentionPancreas} to refine the focus on relevant spatial features at each layer, improving the model’s ability to discriminate between noisy and noise-free regions.
To include an uncertainty quantification component \cite{Cheng2023MachineReview}we introudce a randomized parameterization of observation error covariance $\mathbf{R}$ where  
\begin{equation}
\label{eq:covariance}
\mathbf{R} = \epsilon \mathbf{I} ,
\end{equation}
where, \(\epsilon\) is a scalar Gaussian random variable with mean \(\mu\) and variance \(\sigma^2\). This random parameterization of the observation error covariance effectively varies the balance between the observation term and the prior term in our global minimization problem. After testing on the validation dataset (not shown), we made ensemble generations setting \(\mu = 0.5\) and \(\sigma = 0.1\), corresponding to a standard deviation of 20\% of the mean.

\section{Neural metric space} 
\label{section:neural_metric_space}
Through the mapping operator \(M_{\Phi}\), we aim to learn a neural metric space to evaluate whether a given field aligns with the statistical characteristics of a reference dataset, such as its probability density or higher-order moments. Given the multi-scale nature of geophysical dynamics, we leverage a scattering transform approach readily computable in Python through the \texttt{kymatio} library~\cite{Andreux2020Kymatio:Python}. However, after initial tests, we opted to retain only the first-order wavelet coefficients in our analysis for the sake of interpretability.

Let us denote by \(\phi_j^{(l)}(\mathbf{x})\) the wavelet transform of a given ocean field~\(\mathbf{x}\) at scale \(2^j\) pixels and orientation \(\frac{l}{L} \pi\) radians with $l\in\mathbb{Z}: l\in[1,L]$. In our configuration, we use three different scales, corresponding to spatial resolutions of \(4\,\text{km}\), \(8\,\text{km}\), and \(16\,\text{km}\), and \(L = 8\) orientations spanning \(0\) to \(\pi\). Input data are splited into \(26 \times 26\)-pixel patches (approximately \(52\,\text{km} \times 52\,\text{km}\)). This size was chosen to align with the specifics of our problem, as the SWOT swath is approximately \(120\,\text{km}\) wide with a \(20\,\text{km}\) nadir gap. However, other patch sizes could be selected depending on the application. Next, we pass these wavelet coefficients through an encoder-decoder (autoencoder) architecture to obtain a compact representation. Specifically, we introduce an encoder \(\mathrm{Enc}(\cdot; \mathbf{\Phi}_{\mathrm{enc}})\) and a decoder \(\mathrm{Dec}(\cdot; \mathbf{\Phi}_{\mathrm{dec}})\), with learnable weights \(\mathbf{\Phi}_{\mathrm{enc}}\) and \(\mathbf{\Phi}_{\mathrm{dec}}\). The encoded latent representation \(\mathbf{z}\) is given by
\begin{equation}
\label{eq:encoder}
  \mathbf{z} \;=\; \mathrm{Enc}\!\Big(\big\{\phi_j^{(l)}(\mathbf{x})\big\}_{j,l};\, \mathbf{\Phi}_{\mathrm{enc}}\Big),
\end{equation}
and the reconstructed wavelet coefficients \(\hat{\phi}_j^{(l)}(\mathbf{x})\) are generated by
\begin{equation}
  \hat{\phi}_j^{(l)}(\mathbf{x}) \;=\; \mathrm{Dec}\!\Big(\mathbf{z};\, \mathbf{\Phi}_{\mathrm{dec}}\Big).
\end{equation}
We then define our mapping operator \(M_\Phi\) as the composition of the wavelet encoder defined in \autoref{eq:encoder} and a simple feedforward neural network that aims to reduce the channels of \(\mathbf{z}\) to a single scalar value.

We further impose a sparsity-inducing regularization on the encoder weights via an additional term in the autoencoder training loss:
\begin{equation}
  \mathcal{J}_{\text{AE}} 
  \;=\; \sum_{j, l} \big\|\,\phi_j^{(l)}(\mathbf{x}) - \hat{\phi}_j^{(l)}(\mathbf{x})\big\|_2^2 
  \;+\; \lambda\, \big\|\,\mathbf{\Phi}_{\text{enc}}\big\|_1,
  \label{eq:loss}
\end{equation}
where \(\lambda\) is a hyperparameter controlling the L1-based sparsity constraint on \(\mathbf{\Phi}_{\text{enc}}\). After testing with the validation dataset, we found that \(\lambda = \tfrac12\) is a good compromise. While changing \(\lambda\) primarily affects the number of epochs required for each loss component to converge, it does not hamper the training.  By minimizing~\eqref{eq:loss}, we ensure that the learned feature space retains geophysically-meaningful structure, while also promoting disentanglement in a manner inspired by interpretability approaches in large language models (LLMs)\cite{Cunningham2023SparseModels}.

\section{Neural Metric Characterization of Noise}

In this section, we focus on the neural metric of SIMPGEN. This network is designed to quantify the similarity between oceanographic fields and a given reference dataset. It assigns to each field a numerical dimensionless value that reflects how much this field could belong to the reference. 

During training, the neural metric learns not only from the reference dataset but also from negatively labeled examples, allowing it to differentiate between realistic oceanic fields and those deviating from expected physical structures. In this context, training the network by combining model-based reference fields with real SWOT observations is particularly valuable. This approach enables us to characterize the actual KaRIn noise in real SWOT observations and further examine how simulated SWOT noise differs from real SWOT noise. 

\begin{figure}[ht]
    \centering
    \includegraphics[width=1.0\textwidth]{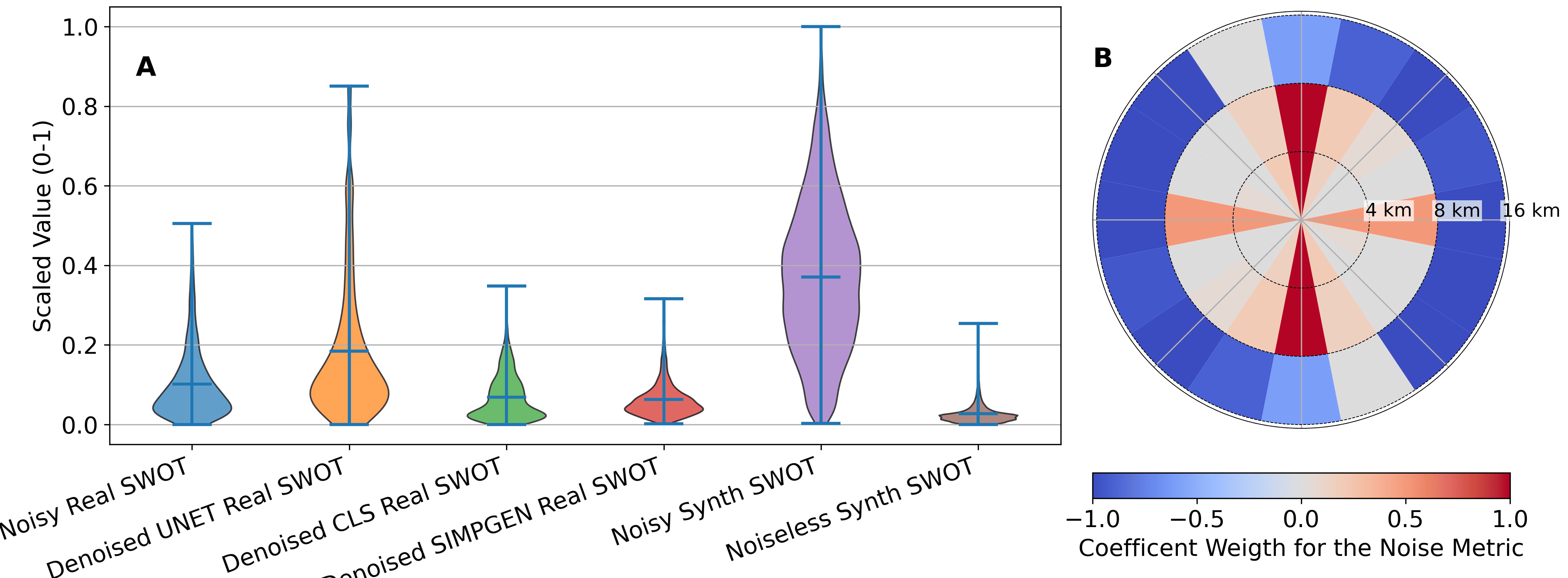}
    \caption{Panels A displays violin plot of the neural metric values for the different dataset. Panel B shows a polar plot of the neural metric sparse autoencoder activations, with positive (red) and negative (blue) values distributed across spatial directions and scales.}\label{fig:histogram_noise_metric}
\end{figure}

\begin{figure}[ht]
    \centering
    \includegraphics[width=1.0\textwidth]{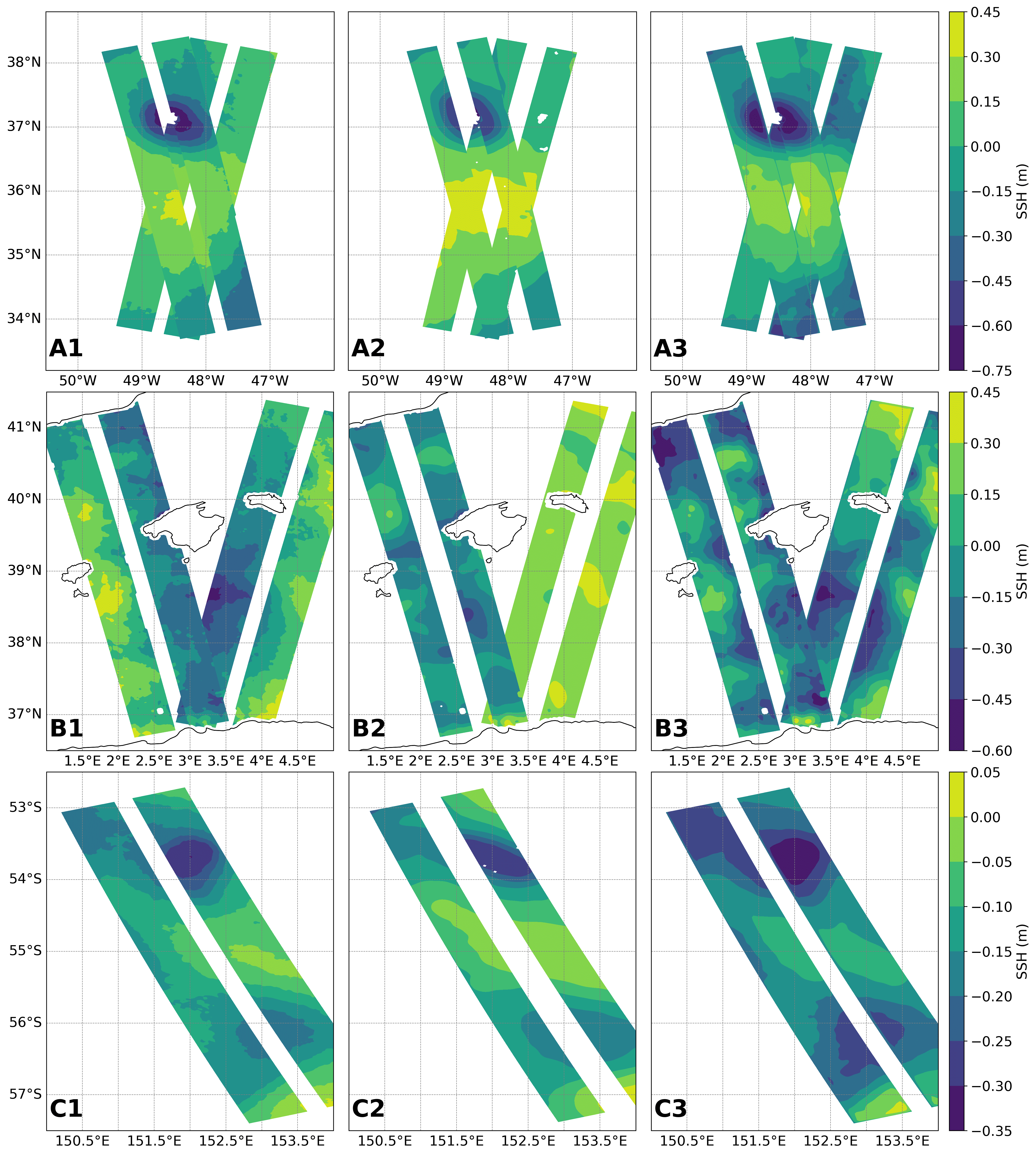}
    \caption{This figure presents a snapshot from 15 April 2023 for the same three regions analyzed in the main paper: the Gulf Stream (A1–A3), the Mediterranean Sea (B1–B3), and the Southern Ocean (C1–C3). The first column shows the raw SSH from SWOT, the second column displays the CLS-denoised SSH, and the third column presents the SIMPGEN-denoised SSH.}\label{fig:ex1}
\end{figure}

\begin{figure}[ht]
    \centering
    \includegraphics[width=1.0\textwidth]{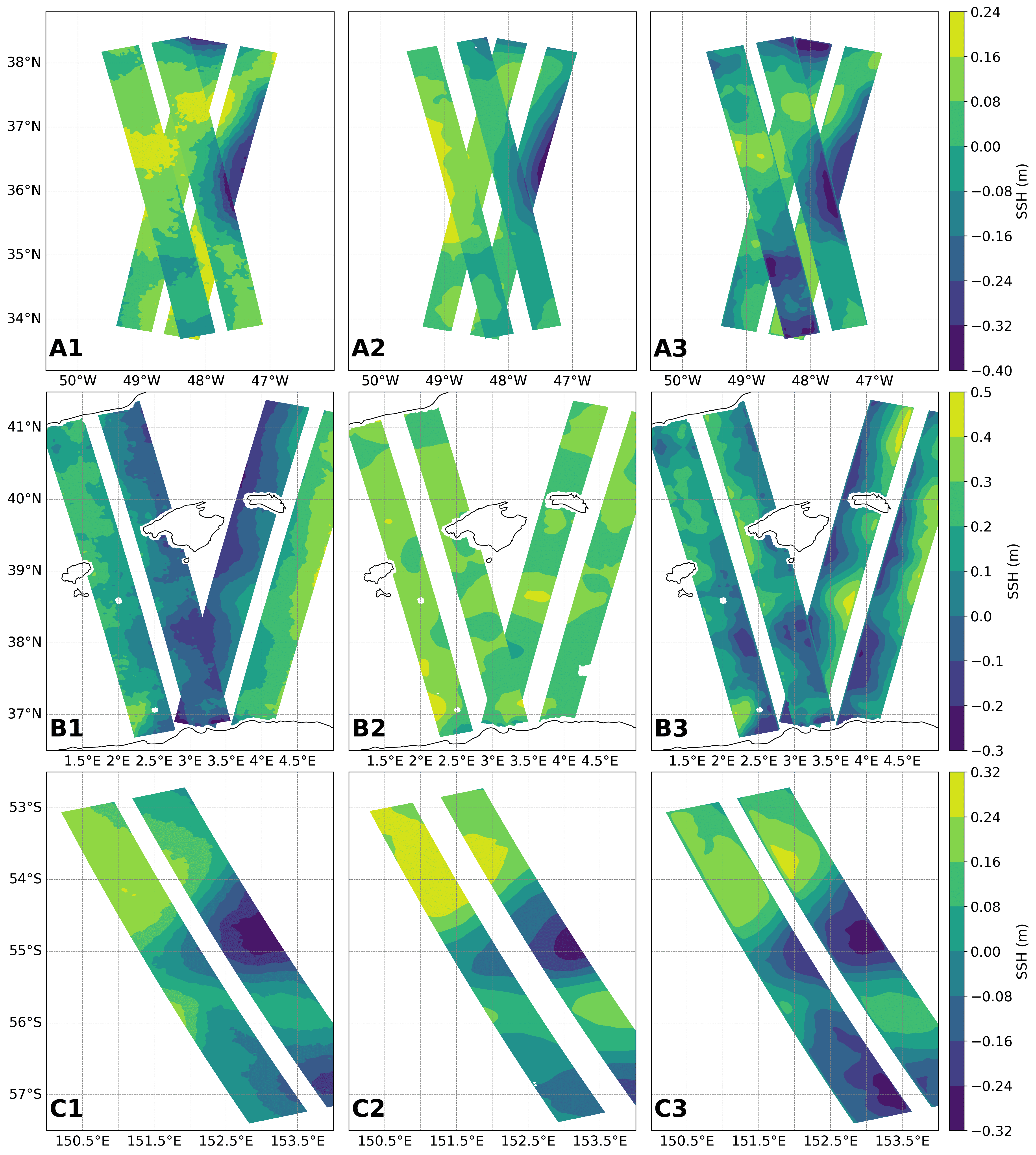}
    \caption{This figure presents a snapshot from 28 June 2023 for the same three regions analyzed in the main paper: the Gulf Stream (A1–A3), the Mediterranean Sea (B1–B3), and the Southern Ocean (C1–C3). The first column shows the raw SSH from SWOT, the second column displays the CLS-denoised SSH, and the third column presents the SIMPGEN-denoised SSH.}\label{fig:ex2}
\end{figure}

In \autoref{fig:histogram_noise_metric}, Panel A, violin plots illustrate the distribution of the neural metric across all datasets. The ‘Noiseless Synth SWOT’ data (far right) being the reference dataset remain closely centered near zero. In contrast, ‘Noisy Synth SWOT’ and especially ‘Noisy Real SWOT’ exhibit wider spreads, reflecting higher variability and underscoring the distinct noise characteristics present in real observations versus purely simulated data. Among the denoised versions (UNET, CLS, and SIMPGEN), CLS provides a noticeable reduction in noise compared to the raw data but still retains a broader distribution than SIMPGEN, which further narrows the spread and lowers the median. This suggests SIMPGEN handles both the overall noise amplitude and variability more effectively than CLS, especially for real SWOT data. Thus, while each method improves data quality, SIMPGEN achieves more substantial noise suppression and consistency in comparison to CLS. These differences, consistent with the spectra presented in the main paper, can also be observed in the six example snapshots (two per studied area) shown in \autoref{fig:ex1} and \autoref{fig:ex2}.

Further insights into the behavior of the neural metric can be gained by examining the coefficients of the sparse autoencoder used within the metric computation. These coefficients indicate how different wavelet components of an input SSH field contribute to its noise metric value, effectively highlighting the dominant patterns that distinguish noisy and clean data. Panel B of \autoref{fig:histogram_noise_metric} shows these learned coefficient weights, where positive weights (red) enhance the noise metric, and negative weights (blue) suppress it. The observed structure shows that these weights are aligned along orthogonal directions, suggesting that the neural metric is particularly sensitive to energy spread across perpendicular wavelet components. This pattern implies that in noiseless SSH fields, energy is more directionally coherent, meaning it is concentrated along specific dominant directions and scales, reflecting the underlying physical structures of ocean dynamics.

Given these results, we can revisit the wavelet polar plots in Figure 3 of the main paper to analyze the impact of denoising on the directional energy distribution. The reduction of energy in the direction orthogonal to the dominant one confirms that noise tends to be spread across multiple, often inconsistent, directions, particularly at smaller scales around 4-8 km.

Moreover, the weight distribution of the neural metric, which places greater emphasis on the across-track direction, along with the fact that across-track dominated snapshots exhibit the most significant changes after denoising, further supports the hypothesis that KaRIn noise is predominantly across-track dependent. This directional bias in noise aligns with the expected instrument-induced errors, reinforcing the effectiveness of the denoising approach in mitigating SWOT-specific artifacts.

\end{appendix}

\end{document}